\documentclass[aps,pra,twocolumn,showpacs]{revtex4-1}%
\usepackage{amsfonts}
\usepackage{amsmath}
\usepackage{amssymb}
\usepackage{graphicx}
\usepackage{dcolumn}
\usepackage{bm}

\begin{document}

\title{Vortex cores in narrow thin-film strips}
\author{V. G. Kogan}
\affiliation{
Ames Laboratory - US Department of Energy and Iowa State University, Ames, Iowa 50011, USA }
\author{M. Ichioka}
\affiliation{Department of Physics, RIIS, Okayama University, Okayama 700-8530, Japan}
\date{\today}

\begin{abstract}
We study  vortex current distributions in narrow thin-film superconducting strips. If one defines the vortex core ``boundary" as a curve where the current reaches the depairing value,  intriguing features emerge. Our conclusions based on the London approach  have only qualitative relevance since the approach breaks down near the core.  Still, the main observation which might be useful  is that the core size near the strip edges is smaller than in the rest of the strip. If so, the Bardeen-Stephen flux-flow resistivity should be reduced near the edges. Moreover, at elevated temperatures, when the depairing current is small, the vortex core may extend to the whole strip width, thus turning into an edge-to-edge phase-slip line. 
  \end{abstract}
\date{\today}
 \maketitle

\section{Introduction}

Long   thin-film strips are essential elements of various superconducting circuits, the current carrying properties of which are determined by vortices residing there or crossing strips and causing non-zero voltages and energy dissipation, see e.g. \cite{Karl}. Many vortex effects can be described by studying the supercurrent distributions, which away of vortex cores  are well represented by the London theory. However, within this theory, vortex cores are treated just as point singularities despite the fact that properties of cores and, in particular, their size and shape are relevant for evaluation the vortex self-energy and its dynamic properties, the flux-flow resistivity is just an  example. 
   
To describe properly the core structure is  challenging even in uniform bulk materials far from sample boundaries.   The question then whether one can extract some information about core shapes from the London current distribution.
For a single vortex in isotropic  bulk superconductors  the current density distribution out of   core is given by $j=(c\phi_0/32\pi^2 \lambda^3)K_1(r/\lambda)$, where $\phi_0$ is the flux quantum, $\lambda$ is the penetration depth, and $K_1$ is the Modified Bessel function. At short distances $r\ll\lambda$, the London current $j= c\phi_0/32\pi^2 \lambda^2r $  reaches the depairing value (defined as $j_d= c\phi_0/16\pi^2 \lambda^2\xi$ near the critical temperature $T_c$) at $r\sim\xi $ with $\xi$ being the coherence length. In fact, this is one of popular ways to define the vortex core size.    ``Sweeping under the rug"  complexities of the vortex core physics, the simple core model  as the circular normal state region of radius $\xi$ provides correct estimates of the core energy as the condensation energy within the core, $\pi\xi^2(H_c^2/8\pi)$, and of the flux-flow resistivity  as the normal resistance within the core, $\rho_f\approx \rho_n 2\pi\xi^2B/\phi_0$ ($\rho_n$ is the normal resistivity,  $B$ is the magnetic induction) \cite{Bard-Stev,Tinkham}.
   
   In thin films, the vortex current distribution differs substantially from the bulk since the stray fields out of the film affect the currents in the film \cite{Pearl}. Evaluation  of these distributions are difficult in particular    in finite samples, where the film edges may cause  drastic modifications of currents for vortices situated at distances of the order of Pearl length $\Lambda=2\lambda^2/d$ from the edges. Exceptions are the small samples, e.g.,  narrow thin-film bridges of a width $W\ll\Lambda$ where effects of  self-fields can be disregarded \cite{75}. Following the above qualitative prescription for determination of the core shape and size, one may expect  different from circular core shapes for vortices in such bridges, the subject of our discussion below.

 \section{Narrow thin-film strips   }

Consider a thin-film strip in the plane $(x,y)$, $x$ axis  is directed across the strip, $0<x<W$, whereas $y$ is along it.     
  For a vortex at $x=a$, $y=0$, the London equations for the film interior read
\begin{equation}
{\bm h}+4\pi\lambda^2 {\rm curl} {\bm j}/c=\phi_0 \,{\hat {\bm z}}\,\delta 
(x-a,y) \,,
\label{e1}
\end{equation}
$\bm h$ is the magnetic field and $\bm j$ is the supercurrent density. 
Averaging this over the thickness $d$, one obtains
\begin{equation}
h_z+2\pi\Lambda\,  {\rm curl}_z {\bf g}/c=\phi_0 \delta ({\bf r}-{\bf a})\,, 
\label{e2}
\end{equation}
where ${\bf g}({\bf r})$ is the sheet current density, ${\bf r}=(x,y)$, 
${\bf a}=(a,0)$, and $\Lambda =2\lambda^2/d$. Other components of Eq.\,(\ref{e1}) turn identities after averaging.
  
In strips of a width $W\ll\Lambda$, the self-field of the current $\bm g$, given by the Biot-Savart integral, is of the order $g/c$, whereas the second term on the left-hand side of
Eq.\,(\ref{e2})  is of the order $g\Lambda/cW$. Hence, the self-field $h_z$ can
be disregarded, unlike the applied field if it exists.
 
It is convenient to introduce a scalar ``steam function" $G({\bm r})$ such that
${\bm g}={\rm curl}\,G{\hat {\bm z}}$ \cite{75}, 
\begin{equation}
 g_x=\partial_yG \,,\,\,\, g_y=-\partial_xG\,. 
\label{e5}
\end{equation}
Then, we obtain for $G$:
\begin{equation}
\nabla^2G=-(c\phi_0/2\pi\Lambda )\delta({\bf r}-{\bf a})\,. 
\label{eq4}
\end{equation}
The boundary conditions $g_x=0$ at the strip edges $x=0,W$  give  $G=0$ at the edges. Thus, the problem is equivalent to that of two-dimensional electrostatic potential of a linear ``charge" $q=c\phi_0/8\pi^2\Lambda $ situated at ${\bm r}={\bm a}$ between two grounded metal plates parallel to $zy$ at $x=0$ and $x=W$. The  solution of Eq.\,(\ref{eq4}) is obtained by conformal mapping \cite{Morse}:
\FL
\begin{equation}
{\rm tanh}\frac{G}{2q} = \frac{{\rm sin}(\pi a/W)\, {\rm sin}(\pi x/W)}{{\rm cosh}(\pi y/W)-{\rm cos}(\pi a/W)\,{\rm cos}(\pi x/W)}\,.
\label{G}
\end{equation}
The alternative way to present this result as due to an infinite sum over $\pm$ vortex images    out of the strip, see e.g. Ref.\,\onlinecite{Sasha},  is equivalent to Eq.\,(\ref{G}), but having   closed form, Eq.\,(\ref{G}) is more convenient for numerical and analytic  work.
 
 Having the stream function $G$, one evaluates the sheet current components of Eq.\,(\ref{e5}) (Mathematica is helpful in this tedious calculation):
  \widetext
          \begin{eqnarray}
{\cal J}_x&=& g_x  \frac{2\pi \Lambda W}{c\phi_0}=-\frac{ \sin  \pi a  \sin \pi x \sinh \pi y }{\cos 2\pi a +\cos 2\pi x  -4\cos \pi a \cos  \pi x \cosh  \pi y + 2 \cosh^2 \pi y   }\,,
 \nonumber\\
 {\cal J}_y&=&g_y \frac{2\pi \Lambda W}{c\phi_0}=\frac{ \sin \pi  a \, (\cos \pi  a -\cos \pi  x \cosh \pi y )}{\cos 2\pi a +\cos 2\pi x  -4\cos \pi a \cos  \pi x \cosh  \pi y + 2 \cosh^2 \pi y   }\,.
 \label{g_xy}
          \end{eqnarray}
          \endwidetext
 \noindent Here ${\cal \bm J}$  is the dimensionless sheet current density in units of $  c\phi_0/2\pi \Lambda W$ and  $x,y,a$ are measured in units of $W$.       
 
\begin{figure}[h]
\begin{center}
\includegraphics[width=7.cm] {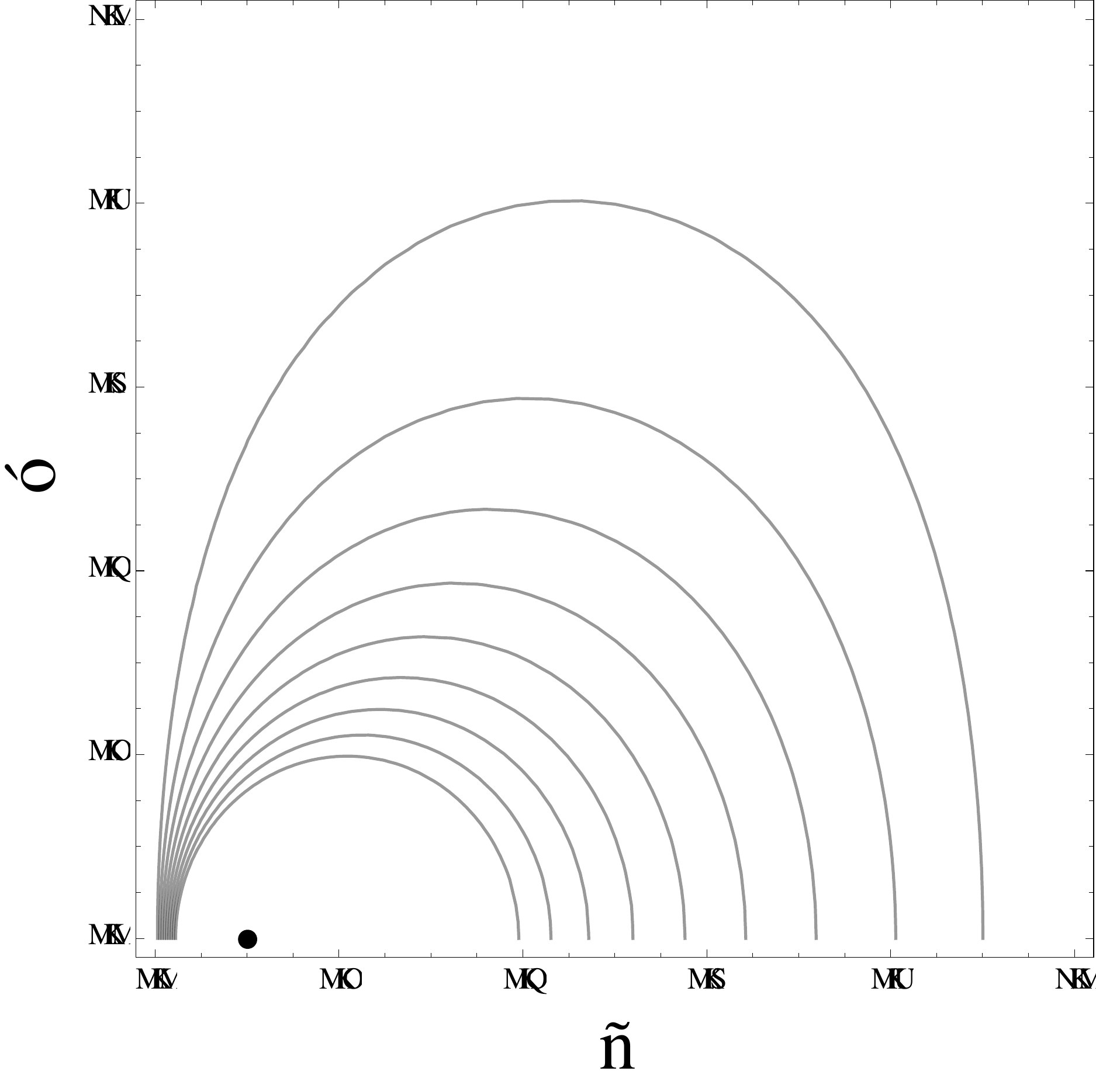}
\caption{ (Color online) Current lines   for a vortex at $a/W=0.1$.  $x,y$ are measured in units of $W$. The black dot marks the position of the vortex singularity. 
}
\label{f1}
\end{center}
\end{figure}

Stream lines of the current coincide with contours of $G=\,$const; indeed,   $(d\bm s\times {  \bm g} )_z =   g_ydx - g_xdy  = dG=0$, where $d\bm s$ is the line element. 
Example of  current stream lines   is shown in Fig.\,\ref{f1} for a vortex close to the left edge of the strip.
  
 To have a better idea on the distribution of current  {\it values},  one can plot  contours of  constant ${\cal J}(x,y) =\sqrt{{\cal J}_x^2+{\cal J}_y^2}$. This distribution near the vortex core  affects the core shape. 
    \begin{figure}[htb]
\begin{center}
\includegraphics[width=8cm] {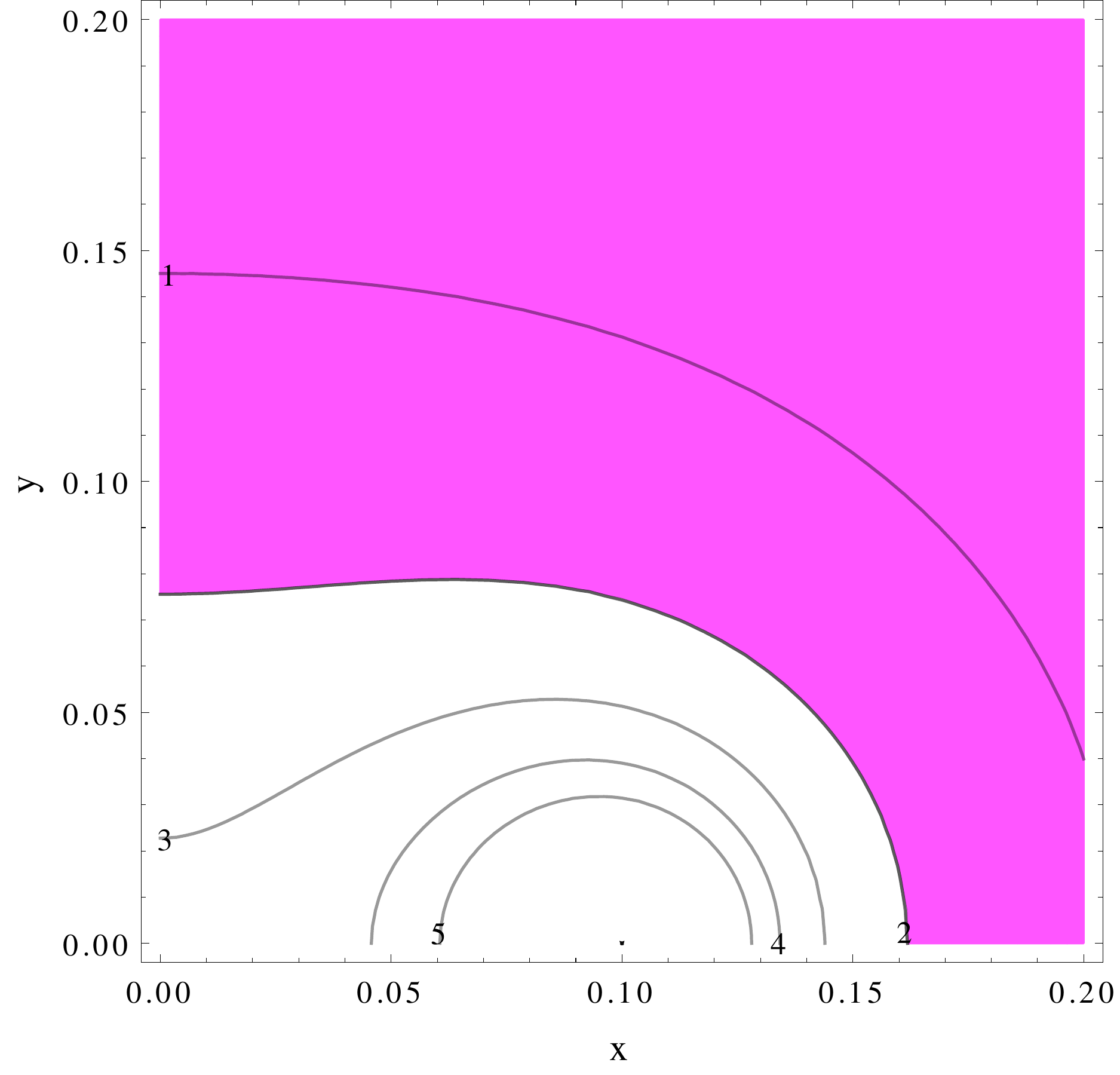}
\caption{ (Color online) Contours of constant current {\it values} for vortex at $a/W=0.1$. The   curve separating the white ``normal core"  from the magenta superconductor  is the estimated ``core boundary" for WSi thin-film strip at  $T/T_c\approx 0.89$  with the depairing current value ${\cal J}_d=2$. The numbers by the contours are ${\cal J}$ values. 
}
\label{f2}
\end{center}
\end{figure}
 Using the standard estimate for the deparing current density one gets the sheet deparing density $g_d\approx c\phi_0d/16\pi^2\lambda^2\xi = c\phi_0/8\pi^2\Lambda \xi$ and  the dimensionless depairing current
\begin{equation}
{\cal J}_d\approx  W/4\pi \xi  \,.
\label{Jd}
\end{equation}
It is worth noting that the depairing current value adopted here is not a universal quantity which may vary (slightly) with the sample geometry. This uncertainty   may introduce an extra factor $\sim 1$ in Eq.\,(\ref{Jd}). 

 If one takes data from Ref. \cite{Andreas} for  WSi    4\,nm-thick narrow bridges with $T_c=3.4\,$K,   the low temperature value  $\xi_0\approx 7.8\,$nm, and    $W=2\,\mu$m, one  estimates the low temperature ${\cal J}_d(0)\approx 20.4$. Hereafter we use these  data  to compare with our model predictions.
 
 On warming, the depairing current decreases  according to empirical relation ${\cal J}_d= {\cal J}_d(0)(1-t^2)^{3/2}$, $t=T/T_c$ \cite{Bardeen,Kunchur}. For our example, ${\cal J}_d\approx 2$ at $T\approx 3\,$K. 
  The   contour ${\cal J}(x,y)=2$ is shown in Fig.\,\ref{f2} for a vortex at $a=0.1 \,W$.  Hence, the core shape, defined by ${\cal J}(x,y)={\cal J}_d$  for a vortex penetrating the edge $x=0$, has the shape of a liquid droplet  stuck to the film edge. This unusual  shape can be attributed to enhanced current density between the edge and the position of the current singularity for a vortex close to the edge   seen in Fig.\,\ref{f1}.
It is worth noting that since ${\cal J}_d$ decreases on warming, according to Fig.\,\ref{f2}   the normal core expands on warming  as it should. 

\subsection{Core shape dependence on  vortex position}

\begin{figure}[ h]
\begin{center}
\includegraphics[width=7.cm] {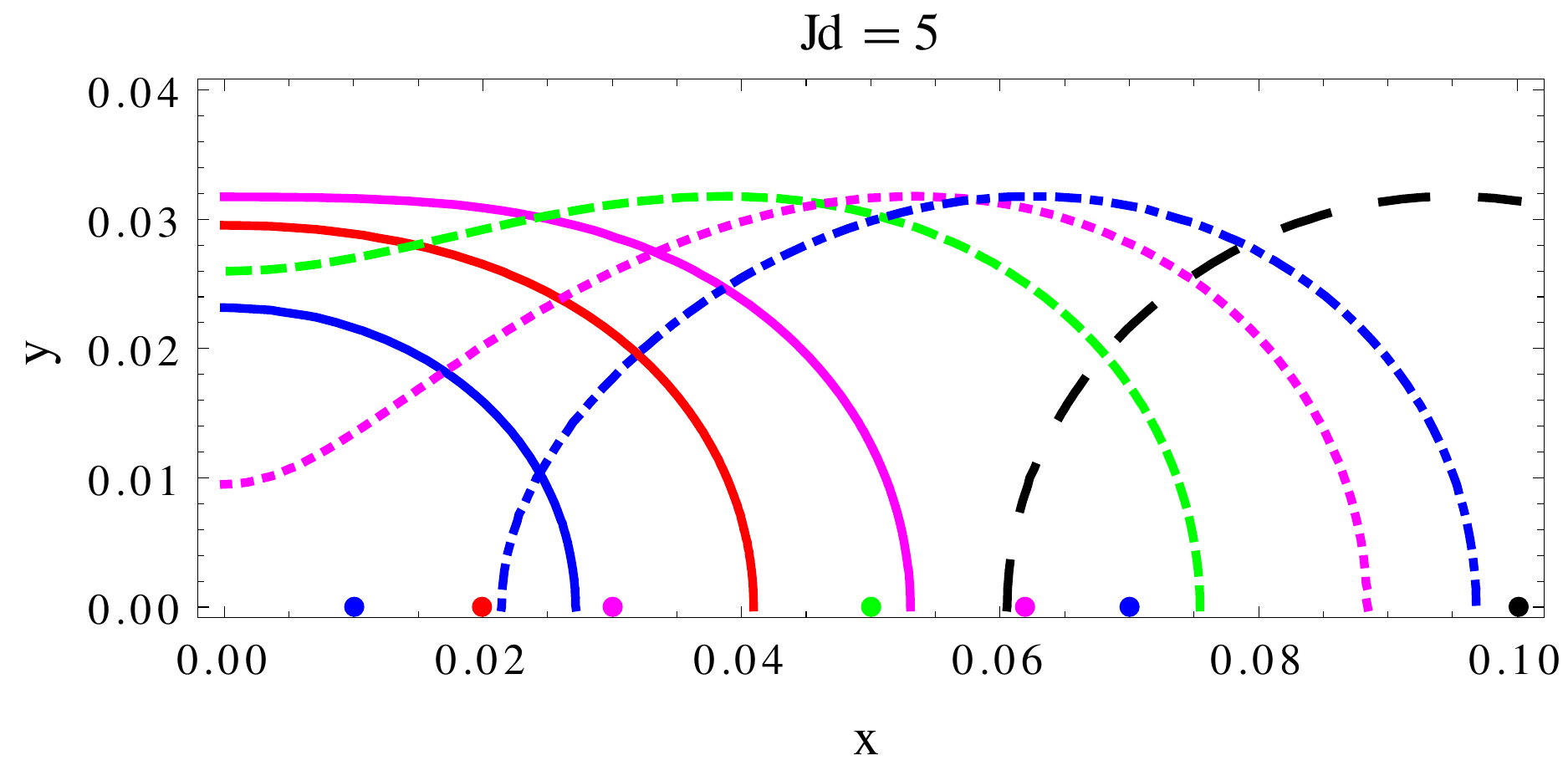}
\caption{ (Color online) Lines of constant current values ${\cal J}(x,y)={\cal J}_d=5$ for vortices at a set of positions $a/W $ marked by   dots: 0.01 (solid blue), 0.02 (solid red), 0.03 (solid  magenta), 0.05 (dashed green), 0.062 (dotted magenta), 0.07 (dash-dotted blue), and 0.10 (dashed black). All curves are symmetric relative to $y=0$, so that the part $y<0$ is not shown. 
}
\label{f3}
\end{center}
\end{figure}

For the above example of thin and narrow bridge of WSi,  $ {\cal J}_d= 5$ corresponds to   $T\approx 0.78\, T_c$. This value of $ {\cal J}_d $ is chosen here to demonstrate the dynamics of the core shape of a vortex moving away from the edge at $x=0$. 
Figure \ref{f3} shows contours $ {\cal J}(x,y)  = {\cal J}_d=5$  for a set of vortex positions near the edge. 

One sees, that  when $a\lesssim 0.03$, the core shape is close to  semi-circles or ovals with a base at the edge, i.e. at the $y$ axis, and with the size growing with increasing $a$. When the distance  $a$ from the edge increases further, the core acquires a shape   reminiscent of a liquid droplet still attached to the edge, see curves for $a=0.05, 0.062$. Eventually, the core droplet disengages from the edge and acquires a nearly circular shape,   $a=0.07, 0.10$. It is readily  shown that in general the disengagement happens at
\begin{equation}
 a=\frac{2}{\pi} \cot^{-1} ( 2 {\cal J}_d ) \,.
 \label{divorse} 
\end{equation}

One also observes that when the vortex proceeds up to  $a\approx 0.03 $, the core expands in both directions. For larger $a=0.05, 0.062$, the increase stops in the $y$ direction along the strip, but the width of the core in $x$ direction  expands.   

\begin{figure}[ h]
\begin{center}
\includegraphics[width=7.5cm] {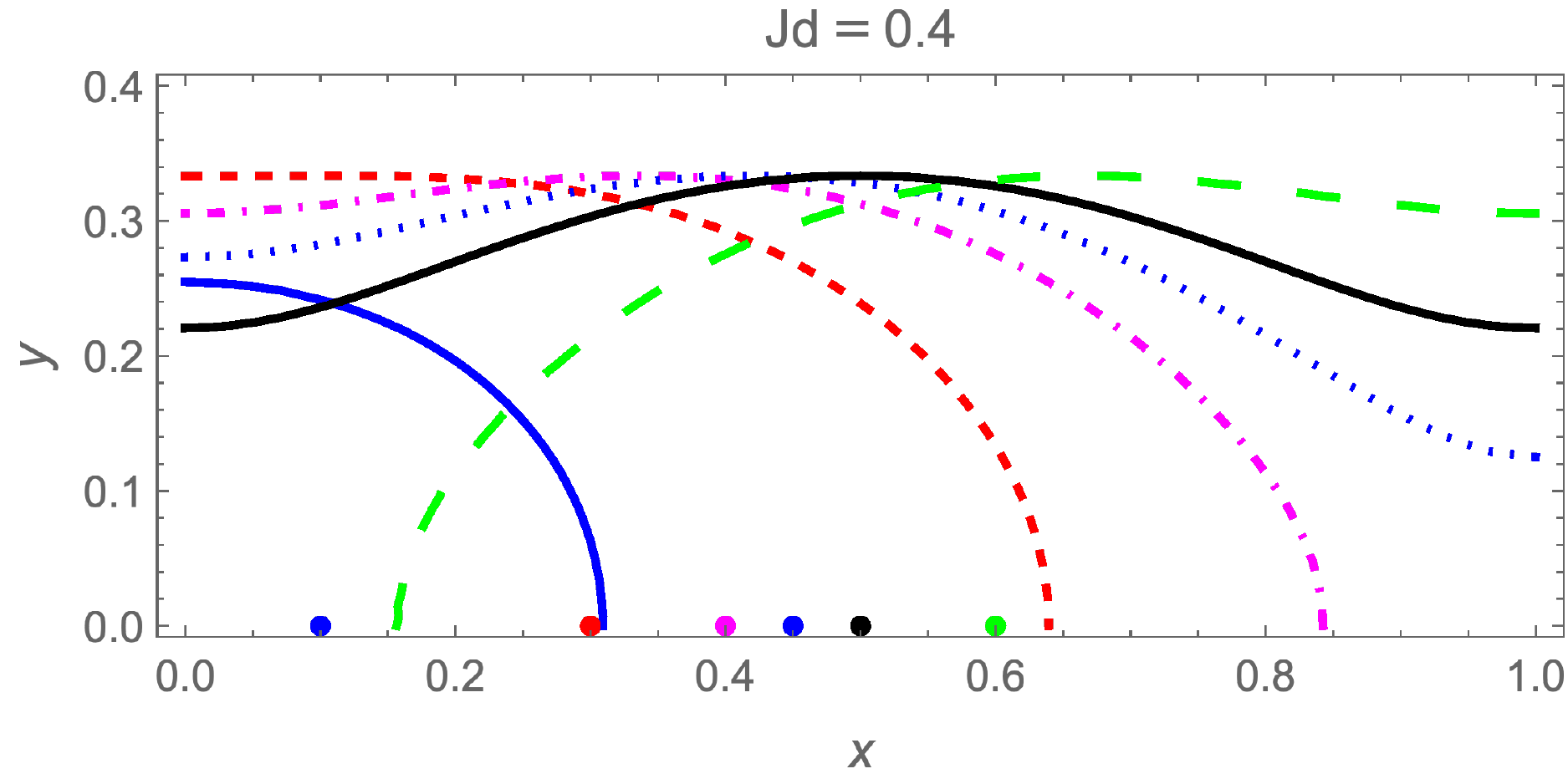}
\caption{ (Color online) Lines of constant current values ${\cal J}(x,y)={\cal J}_d=0.4$ for  vortices at a set of vortex positions $a $   marked by  dots:    0.1 (solid blue),   0.3     (dashed red),   0.4 (dash-dot  magenta), 0.45 (dotted blue), 0.5 (solid black), and 0.6 (dashed green).   
}
\label{f4}
\end{center}
\end{figure}

The behavior of so defined core is even more intriguing at higher temperatures and lower ${\cal J}_d$. An example of ${\cal J}_d=0.4$,  shown in Fig.\,\ref{f4}, corresponds to $T\approx 3.3\,$K$=0.96\,T_c$. One can see that  up to $a=0.4 $ the core, being still attached to the left edge, ends up at some $x^*<1$. One readily verifies  that $x^* $ satisfies 
          \begin{eqnarray}
{\cal J}(x^*,0,a) =\frac{ \sin \pi  a} {2|\cos \pi  a-\cos \pi  x^*|}={\cal J}_d \,.
  \label{J(x0a)}
          \end{eqnarray}

\noindent However, for $a=0.45 $ and 0.5 the core turns into a normal state edge-to-edge channel. One may expect these channels to behave as line-type phase slips \cite{Ustinov}.

 Recall that one-dimensional phase slips are responsible for dissipation in wires thinner than $\xi$ \cite{Tinkham}. When a localized  vortex crosses a  superconducting strip of length $2L$ with the width $W\gg \xi$,   the phase difference between edges $y=-L$  and $L$ also ``slips" by $2\pi$ as in thin wires, see  e.g.  \cite{BGK}. The line phase slips were proposed as another mechanism of dissipation in wide strips \cite{Ustinov}. Hence, one may expect a similar behavior in thin-film strips of interest here. 

    \begin{figure}[t]
\begin{center}
\includegraphics[width=7.cm] {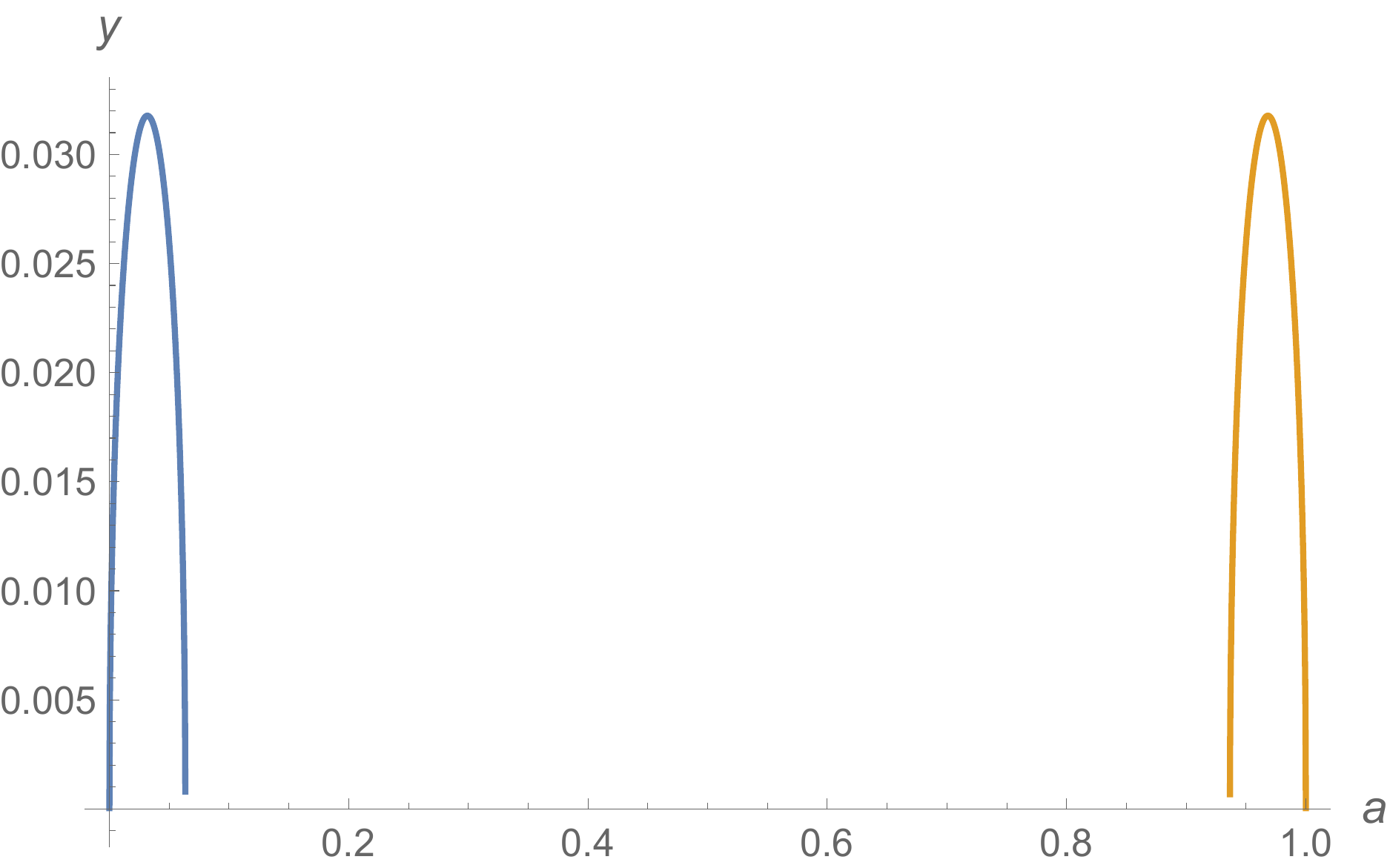}
\includegraphics[width=7.cm] {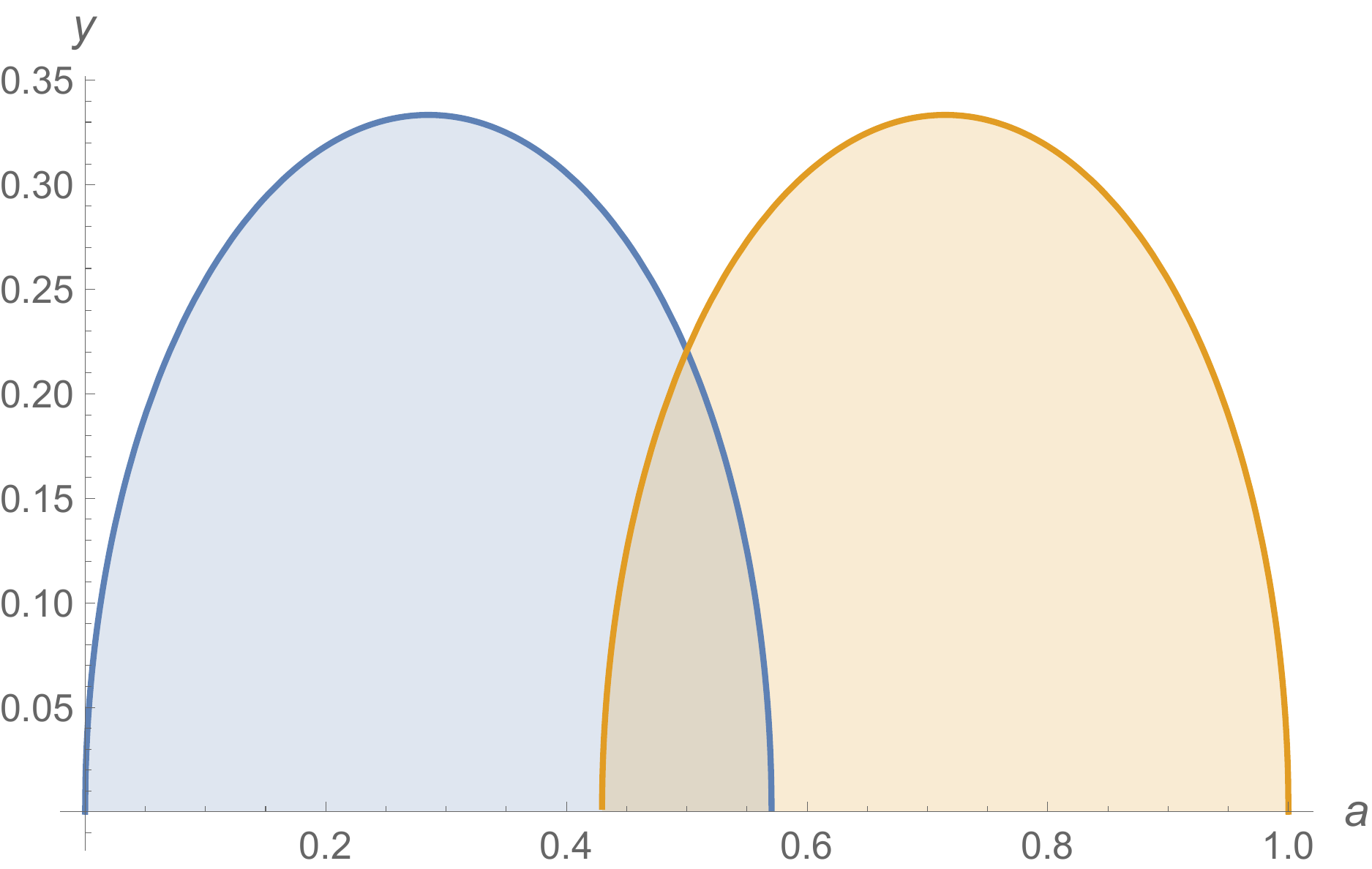}
\caption{ (Color online) The upper panel: the width $y$ of the droplet base at $x=0$ as function of the vortex position $a$ for ${\cal J}_d=5$; the droplet base at the left edge is finite for $0<a\lesssim 0.06$ and at the right edge if $ 0.94\lesssim a<1$.  At other  positions cores are disconnected from both edges. The lower panel: the same for   ${\cal J}_d=0.4$. For all $a$, the core droplet is connected to one of the edges; moreover,  domains of nonzero droplet bases are overlapped if $0.43\lesssim a\lesssim 0.57$, i.e.,   the core occupies an end-to-end belt as in line phase slips. 
}
\label{fig5new}
\end{center}
\end{figure}

One can also look at the disengagement from the edge calculating the width $2 y_1$ of the droplet base at the left edge at $x=0$ where  $y_1$ is found from  $|{\cal J}_y(0,y_1,a)| ={\cal J}_d$. This gives 
          \begin{eqnarray}
\cos\pi y_1 =\frac{ \sin \pi  a} {2 {\cal J}_d}+\cos\pi a \,.
  \label{y1}
          \end{eqnarray}
Similarly, for the edge at $x=W$, one obtains from $|{\cal J}_y(W,y_2,a)| ={\cal J}_d$:
          \begin{eqnarray}
\cos\pi y_2 =\frac{ \sin \pi  a} {2 {\cal J}_d}-\cos\pi a\,.
  \label{y2}
          \end{eqnarray}
          
Fig.\,\ref{fig5new} shows $y_1(a)$ and $y_2(a)$ for ${\cal J}_d=5$ (the upper panel) and ${\cal J}_d=0.4$ (the lower panel). One sees that at low temperatures with large 
${\cal J}_d$  the vortex cores are separated from the edges for most of vortex positions in the  sample, except narrow belts near the edges. These belts expand on warming and at ${\cal J}_d=0.5$, all vortices at $a<0.5$ are stuck to the left edge, while those at $a>0.5$ are  attached to the right edge. With further warming ${\cal J}_d<0.5$, the two domains overlap, as shown at the lower panel, in other words, in a finite interval of positions centered at the sample middle the cores are attached to both edges. That is the situation when the cores are expected to behave as line phase slips.  

\section{Discussion}

A word of caution: 
the definition of the ``core boundary"  as a curve  where the current {\it values} in the London approximation reach  the depairing level is rather artificial, notwithstanding reasonable results  it leads to  in isotropic bulk situation. 
In fact, the London approach breaks down near these boundaries. Hence, a better theory should be employed in the core vicinity. To reaffirm the contours of $J(x,y)=J_d$    in Figs.\,\ref{f2}-\ref{f4} as representing, at least qualitatively,  vortex core shapes, one has to see whether or not the order parameter modulus $|\Delta(x,y)|$ indeed decreases when one crosses these contours. This, of course, cannot be done within the London model where $|\Delta|$ is assumed constant. One should turn to a microscopic theory  or, for temperatures close to  $T_c$,   to the Ginzburg-Landau theory (GL). 

The problem of  the  order parameter distributions for vortices perpendicular to a  thin film, is challenging  because  one has to take the stray fields into account. 
Given these difficulties, one could turn to other geometry, where the comparison between the current and order parameter distributions is easier to make. Such a case is a vortex parallel to faces of a thin superconducting slab   \cite{Sasha}. 

It is straightforward to show that the London current distribution for a vortex parallel to a slab  thinner than the London $\lambda$ are, in fact, the same as in narrow thin-film bridges discussed above, except now we have the length scale $\lambda$ instead of  Pearl's $\Lambda$ and the slab thickness instead of the bridge width $W$. We are not aware of calculations describing the order parameter distribution within the core of such a vortex. 
Physically, however, such a distribution is similar to that of a vortex close to the surface of a bulk sample, the problem considered within the frame of time-dependent GL theory \cite{Kato}, and recently    in the discussion of surface barriers near $T_c$  \cite{Babaev}. Contours of constant order parameter $|\psi |$ shown in Fig.\,1 of Ref.\,\onlinecite{Babaev} are indeed qualitatively similar to  contours of our Fig.\,\ref{f2} of constant current value.


To conclude, we have shown that along with the distorted vortex current distribution near the film edges,     vortex cores  are  strongly affected as well. Close to the edge, the core is shaped as a liquid droplet attached to the edge, which grows when vortex singularity moves away from the edge. At some distance, depending on the depairing current value, the core disconnects from the edge, Eq.\,(\ref{divorse}), and acquires a ``normal" round shape. If  temperatures are high, the depairing current is small, and the thin-film bridge is narrow, the core can be attached to both edges simultaneously thus forming a structure similar to line-type   phase slips. 
  
Our discussion  could be relevant for interpretation of data on resistivity  transition from the normal to superconducting state in narrow thin-film strips reported in  \cite{Andreas}. In particular, these data were shown to be consistent with enhanced flux-flow conductivity  near the strip side edges   in applied  fields on the Tesla order. The conductivity variation could be associated with   changing core size of vortices crossing the strips. \\

The work of V.K. was supported by the U.S. Department of Energy, Office of Science, Basic Energy Sciences, Materials Science and Engineering Division.  Ames Laboratory  is operated for the U.S. DOE by Iowa State University under contract \# DE-AC02-07CH11358. 
The work of M.I. was supported by JSPS KAKENHI Grant No.17K05542.


\end{document}